 \documentclass[final,3p,times,twocolumn]{elsarticle}

\usepackage{amssymb}
\usepackage{lineno}
\usepackage{amsmath,amsfonts}
\usepackage{graphicx}
\usepackage{subfigure}
\usepackage[perc]{overpic} 
\usepackage{gensymb}
\usepackage{xcolor,varwidth}

\journal{Nuclear Physics A}

\begin{document}

\begin{frontmatter}



\title{Evolution of the electric fields induced in high intensity laser-matter interactions}

\author[a]{F.G. Bisesto}
\author[a]{M.P. Anania}
\author[b]{M. Botton}
\author[a]{E. Chiadroni}
\author[c,d]{A. Cianchi}
\author[a]{A. Curcio}
\author[a]{M. Ferrario}
\author[e]{M. Galletti}
\author[b]{Z. Henis}
\author[a]{R. Pompili}
\author[b]{E. Schleifer}
\author[a,b]{A. Zigler}

\address[a]{INFN Laboratori Nazionali di Frascati, Via Enrico Fermi 40, 00044 Frascati, Italy}
\address[b]{Racah Institute of Physics, Hebrew University, 91904 Jerusalem, Israel}
\address[c]{University of Rome "Tor Vergata", Physics Department, Via della Ricerca Scientifica 1, 00133 Roma, Italy}
\address[d]{INFN-Roma Tor Vergata, Via della Ricerca Scientifica 1, 00133 Roma, Italy}
\address[e]{Istituto Superior Tecnico de Lisboa, 1049-001 Lisbon, Portugal}

\begin{abstract}
Multi MeV protons \cite{snavely2000intense} and heavier ions are emitted by thin foils irradiated by high-intensity lasers, due to the huge  accelerating fields, up to several teraelectronvolt per meter, at sub-picosecond timescale \cite{dubois2014target}. The evolution of these huge fields is not well understood till today. Here we report, for the first time, direct and temporally resolved measurements of the electric fields produced by the interaction of a short-pulse high-intensity laser with solid targets. The results, obtained with a sub-$100$ fs temporal diagnostics, show that such fields build-up in few hundreds of femtoseconds and lasts after several picoseconds.
\end{abstract}

\begin{keyword}

High power laser \sep plasma acceleration \sep high brightness electron beam \sep electron diagnostics



\end{keyword}

\end{frontmatter}


\section{Introduction}
Femtosecond lasers with extremly high intensities represent an affordable tool to accelerate particles to relativistic energies within very short distances \cite{leemans2006gev}. The emission of fast ions from plasmas produced at the interactions of ultra-short high-power lasers with solid targets, in particular, has attracted a great interest related to the development of small-scale ion accelerators thanks to the ultra-high accelerating gradients achievable. The generated electric fields scale as \cite{abicht2016tracing} $E_0\approx\sqrt{2*I_L/(\epsilon_0 c)}$, where $I_L$ is the laser intensity, $\epsilon_0$ the vacuum dielectric constant and $c$ the speed of light. Therefore, by using nowadays accessible laser intensities of the order of $10^{18}$ W/cm$^2$, huge fields of the order of $3$ TV/m can be produced, i.e. about four orders of magnitude with respect to state-of-the-art RF accelerators.

The physical picture of the ion-acceleration process is the following.  At the early stages of the interaction, the faster electrons may escape from the volume in which ionization took place \cite{singh2013direct}. After their emission, a positive unbalanced charge is left on target, leading to the formation of a macroscopic electrostatic potential responsible of ion acceleration \cite{macchi2013ion}. The potential translates in an attracting force for electrons, thus the ones that do not have enough energy to escape remain locked at the vicinity of the target surface within a distance of the order of the Debye length. The capacitor-like electric field established between the negative electron sheath and the positive charged target surface is able to strip and accelerate protons and ions from the latter one in the form of ultra-short bursts of picosecond duration \cite{patel2003isochoric}. This is the underlying principle of \emph{Target Normal Sheath Acceleration} (TNSA) \cite{wilks2001energetic}. The whole process, however, strongly depends on the strength and lifetime of the electrostatic potential that, in turn, varies with time since the unbalanced positive charge left on target is gradually neutralized by the electrons coming from outer sections \cite{dubois2014target}.  

A direct experimental evidence of the temporal evolution of the electrostatic potential requires sub-picosecond measurements of charge density near the surface or alternatively tracing down the escaping electrons. So far only nanosecond-resolution measurements or indirect time integrated evidences the of radiated electromagnetic pulses \cite{jackel2010all,nilson2012time} and magnetic fields \cite{sandhu2002laser} have been reported.
Here, we present temporally-resolved measurements related to the quasi-static electric field generated on the target surface after the interaction with an high-power ultra-short laser pulse. The results, obtained with an Electro-Optical Sampling (EOS) temporal diagnostics \cite{bisesto2017innovative,bisesto2017innovative2,pompili2017electro,bisesto2017novel}, show for the first time the temporal evolution of electrostatic potential induced by the unbalanced positive charge left on the target. 

\section{Experimental setup}
\begin{figure}[htb!]
\centering
\includegraphics[width=0.95\columnwidth]{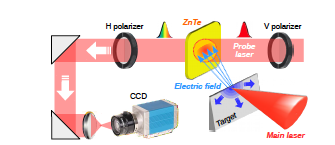}
\caption{Setup of the experiment. The FLAME laser is focused on the tip of a stainless steel blade, producing electron jets that escape from the target and positively charge it. The unbalanced charge (blue region) gradually spreads over the entire target surface, setting up a quasi-static electric field that in turn accelerates ions. It is then sampled by a linearly polarized probe laser while passing through a $500\ \mu$m thick ZnTe electro-optic crystal. The field temporal profile is encoded on the probe laser, resulting in a modulation of its polarization. Finally, a polarizer downstream the crystal converts it into an intensity modulation detected by the CCD camera.}
\label{setup}
\end{figure}

The experiment, depicted in Fig. \ref{setup}, has been performed with the FLAME laser \cite{flame} at the SPARC LAB Test Facility \cite{ferrario2013sparc_lab} by focusing its high-intensity ultra-short-pulses (up to $4$ J energy and $35$ fs pulse duration) on stainless steel solid wedged targets. The EOS employs a $10\times10$ mm$^2$
ZnTe electro-optic crystal with $500\ \mu$m thickness and a $35$ fs (fwhm) probe laser, directly split from the main one to ensure a jitter-free synchronization \cite{pompili2016femtosecond}. The system is able to provide single-shot and non-destructive measurements of the electric field emitted by the target and impinging on the crystal with less than $100$ fs resolution \cite{pompili2016sub}, allowing to operate on the process timescale, determined by the duration of the driving laser pulse \cite{dubois2014target}. A delay-line with $3$ fs resolution is used in order to synchronize the probe and the main lasers in correspondence of the EOS crystal. The probe laser crosses the crystal with $\theta_i=28^{\circ}$ incidence angle, realizing a spatial encoding of the target electric field along the laser probe transverse profile \cite{cavalieri2005clocking}. In such a way the temporal coordinate $t_i$ of the target electric field is related to the laser transverse one $x_i$ by the relation $t_i = x_i \sin{\theta_i} /c$, with $c$ the vacuum speed of light. Being $6$ mm the diameter of the probe laser, the resulting active time window provided by the EOS is approximately $10$ ps. In order to allow the field to freely propagate toward the diagnostics (located at $1$ mm distance), we focused the FLAME laser on the tip of the wedged target, directly looking at the EOS crystal.

\begin{figure}[htb!]
\centering
\includegraphics[width=0.9\columnwidth]{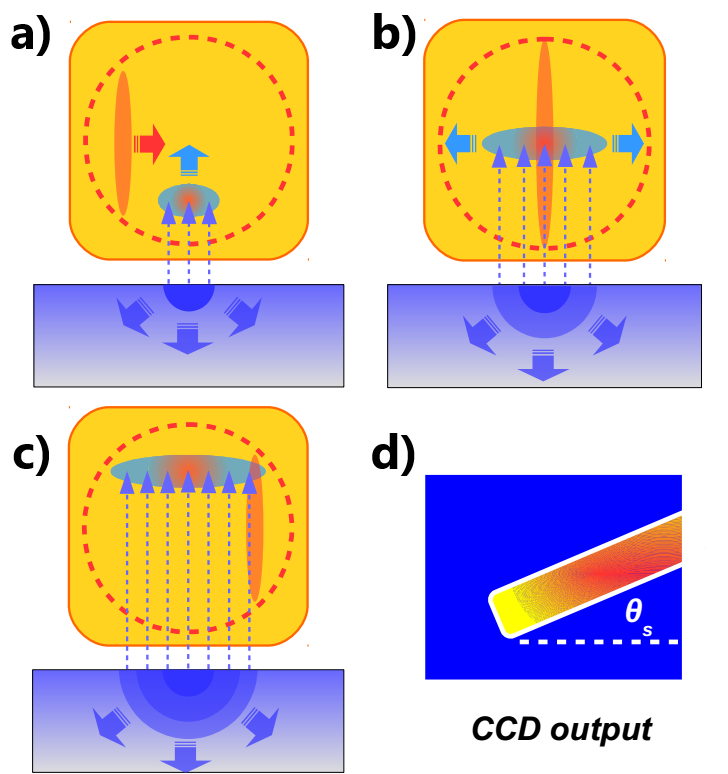}
\caption{Electro-optic encoding. (a) The unbalanced positive charge on target (blue region) generates an electric field (dashed arrows) that impinges on the EOS crystal (yellow square), inducing the electro-optic effect. (b) A linearly polarized probe laser simultaneously enters the crystal with an angle, moving through it from left to right (red ellipse) and spanning an overall region equal to its transverse spot size (red dashed circle). (c) While the charge spreads on target, the electric field source grows in size covering a larger and larger area on the crystal. (d) The probe polarization is locally modulated (red rectangle) as a result of the overlapping between the probe laser and target electric field. From the length and thickness of such modulation, the target field source size and duration can be retrieved.}
\label{encoding}
\end{figure}

The encoding process of the target electric field along the laser probe is sketched in Fig. \ref{encoding}. The unbalanced positive charge (blue semicircle), left on target by the escaped electrons, generates an electric field that propagates in the space around (Fig. \ref{encoding}a). With time, the charge gradually spreads over the target surface, covering a larger and larger area (Fig. \ref{encoding}b). Simultaneously, the linearly polarized probe laser pulse (red ellipse) laterally crosses the crystal and samples the local birefringence (green ellipse) induced by the target field (Fig. \ref{encoding}c). As a result, the probe laser overlaps with the target field along a tilted cigar-like shape (red rectangle) whose thickness and length are proportional to the field duration and transverse size (i.e. the charged area on target that induced the field itself), respectively (Fig. \ref{encoding}d).

\section{Experimental results}
\begin{figure}[htb!]
\centering
\includegraphics[width=0.9\columnwidth]{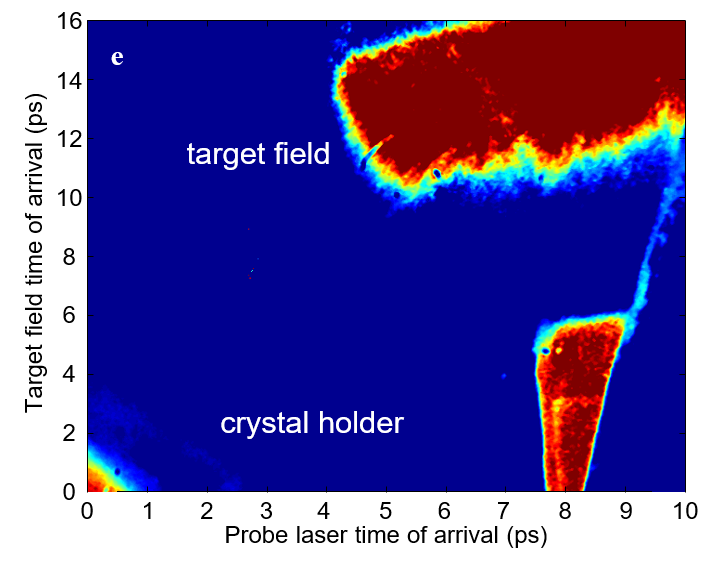}
\caption{Single-shot measurement resulting from the laser-target interaction.}
\label{fig:typicalresults}
\end{figure}

\begin{figure}[htb!]
\centering
\includegraphics[width=0.9\columnwidth]{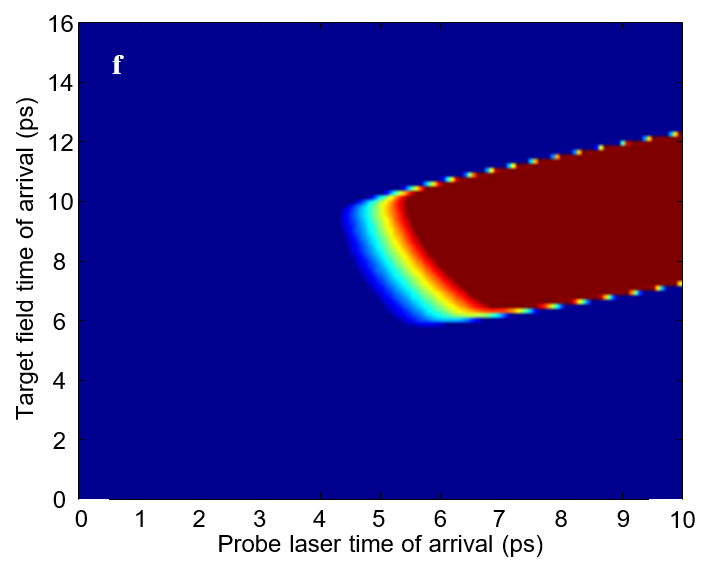}
\caption{Simulation of the electro-optic signal as detected by the CCD.}
\label{fig:simulations}
\end{figure}

Figure \ref{fig:typicalresults} shows a typical single-shot electro-optic snapshot obtained by focusing the FLAME laser on a stainless steel blade target. The laser spot radius on target is $r_L\approx10\ \mu$m, corresponding to an intensity $I_L\approx2\times10^{18}$ W/cm$^2$. The signal clearly exhibits a straight and tilted shape, as expected from the electro-optic spatial encoding. Due to the geometry of the setup (the probe laser crosses the crystal from left to right while the target field moves from bottom to top), the x and y axes represent, respectively, the probe laser and target field time of arrival onto the crystal. In order to fully understand the resulting electro-optic signal shape, we developed a numerical simulation code reproducing the EOS response as detected by the CCD camera. The simulation assumes an initial positive charge concentrated within a circular region of the target with initial radius $r_L$ . The charge gradually spreads in time over the target, with its radius growing approximately at speed of light \cite{dubois2014target}. The so induced electric field propagates in vacuum reaching the EOS crystal and, as a result, it is imprinted along a "cigar-like" shape on the probe transverse profile, like in Fig. \ref{fig:typicalresults}. Figure \ref{fig:simulations} reports the simulated output obtained by assuming $100$ nC charge on target producing an electric field with $6$ ps duration. Both the experimental and simulated signals results in an intensity saturation of the CCD camera.
In order to investigate the generation and evolution of the electrostatic field that develops on target we performed a series of single-shot measurements in different experimental conditions, analyzing in particular the effects produced by changing the energy of the interacting laser. We expect that larger is the energy deposited on target, longer is the duration of the induced electrostatic potential \cite{dubois2014target,poye2015physics}. Being $2$ J the maximum energy per pulse, we examined this conjecture by focusing the FLAME laser on the tip of the metallic target with $10\%$, $50\%$ and $100\%$ energy. 
 
\begin{figure}[htb!]
\centering
\includegraphics[width=0.9\columnwidth]{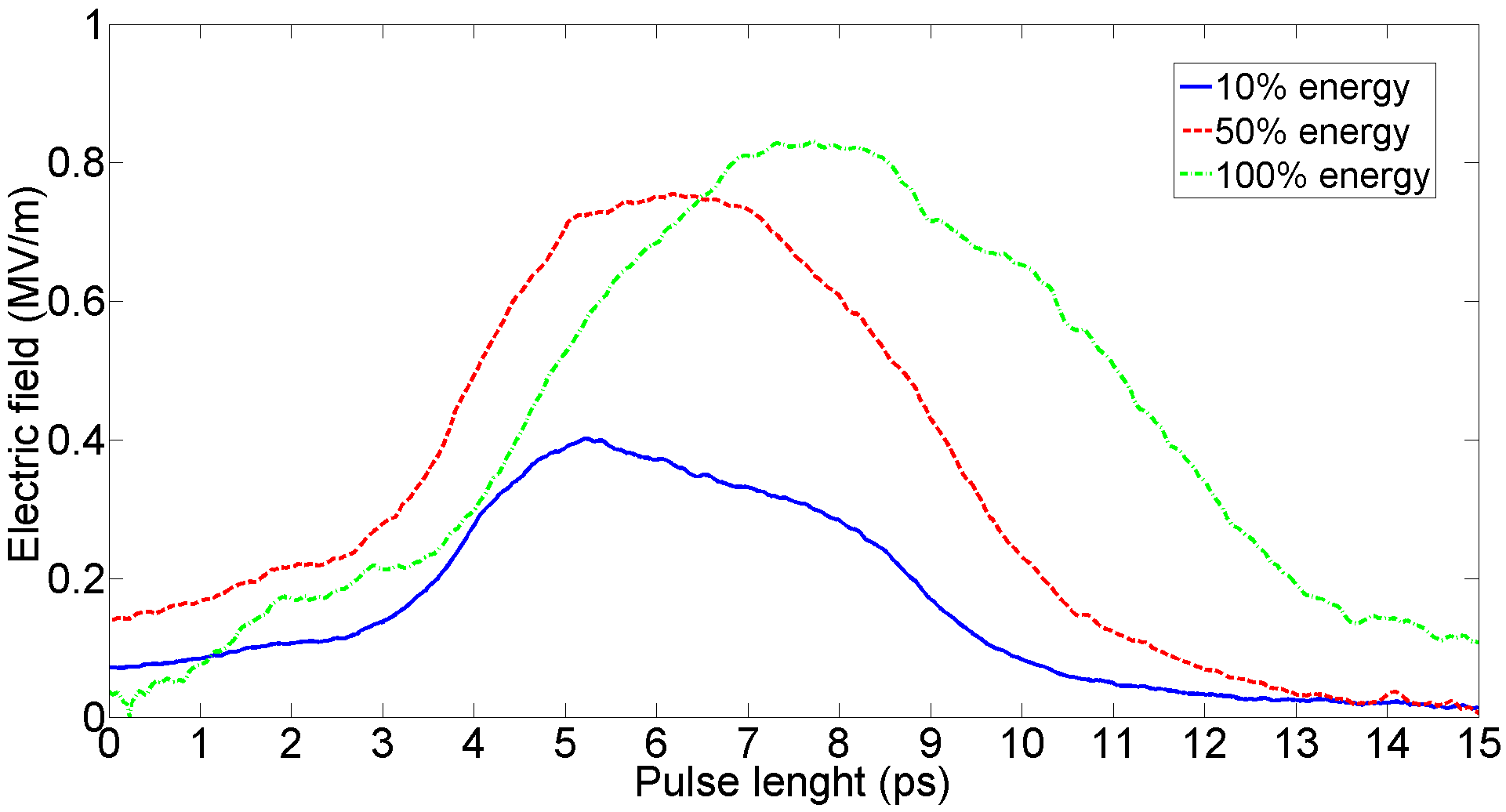}
\caption{\textbf{Detected EM pulses.} The red (green) profile of the radiation pulse $E_T$ is retrieved from the vertical signal thickness at $50\%$ laser energy ($100\%$ energy). The full width half maximum (FWHM) duration is 5.7~ps (7.1~ps). When the laser energy is lowered down to $10\%$ the duration decreases to about 5~ps (blue).} 
\label{eField_cutted}
\end{figure}

By measuring the amount of birefringence induced into the ZnTe crystal, we can extrapolate the amplitude of the electrostatic target field. Since the electrostatic field drops as $r^{−2}$, we have to know the exact distance $r$ between the target and the area of the crystal where the temporal overlapping of the probe laser and the target field is realized. This distance can be estimated by considering that, in all the snapshots acquired during the experiment, the electro-optic signals are emitted from the top-right part of the crystal image plane. We can thus assert that such position corresponds to $r\approx6$ mm distance from the target. The experimental results are shown in Fig. \ref{eField_cutted}, where the line profiles represents the calculated temporal profile of the measured electrostatic target field. When $10\%$ of laser energy is used (blue line), the resulting duration is about $5$ ps (FWHM), while with $50\%$ it is $5.7$ ps (red line). At full laser energy (green line), the potential lifetime increases, resulting approximately $7.1$ ps.
As expected, the electric field has a longer duration when the laser is operated at maximum energy. The field shows also an increased strength due to the fact that a larger amount of electrons have been extracted from target and, in turn, a larger unbalanced positive charge has been left on it.
It is worth pointing out that such signals represent the first measurements ever done of the target electric fields with sub-picosecond resolution, while previous experiments only reported about data obtained on the nanosecond scale \cite{poye2015dynamic}. 
 

\section{Conclusions}
In conclusion we provided femtosecond resolution measurements that allow to operate on the same time scale of
 the considered process, determined by the duration of the driving laser pulse. In our experiment we have revealed the temporal evolution of electromagnetic pulses with lifetime of several picoseconds, emitted from metallic targets after the interaction with an high-intensity short pulse laser. Our study opens the way to perform many new time-resolved experiments with the goal to have a closer and more complete vision of the phenomena involved in laser-matter interactions.

\section*{Acknowledgments}
This work was supported by the European Union‘s Horizon 2020 research and innovation programme under grant agreement No. 653782.

\section*{References}
\bibliographystyle{elsarticle-num} 
\bibliography{eaac17_bib}




%
\end{document}